\documentclass[12pt]{iopart}

\begin{document}

\title[Stationary von Neumann-Landau wave equations]
{Dirichlet problems for stationary von Neumann-Landau wave
equations}

\author{Zeqian Chen}

\address{Wuhan Institute of Physics and Mathematics, Chinese Academy of
Sciences, 30 West District, Xiao-Hong-Shan, P.O.Box 71010, Wuhan,
China}
\ead{zqchen@wipm.ac.cn}
\begin{abstract}
It is known that von Neumann-Landau wave equation can present a
mathematical formalism of motion of quantum mechanics, that is an
extension of Schr\"{o}dinger's wave equation. In this paper, we
concern with the Dirichlet problem of the stationary von
Neumann-Landau wave equation:$$\left \{\begin{array}{l}\left ( -
\triangle_x + \triangle_y \right ) \Phi (x, y) = 0,~~~ x, y \in
\Omega,\\[0.4cm]
\Phi|_{\partial \Omega \times \partial \Omega} = f,
\end{array}\right.$$ where $\Omega$ is a bounded domain in
$\mathbf{R}^n.$ By introducing anti-inner product spaces, we show
the existence and uniqueness of the generalized solution for the
above Dirichlet problem by functional-analytic methods.

\end{abstract}

\maketitle

\section{Introduction}\eqnobysec
In the most general form, Heisenberg's equation and
Schr\"{o}dinger's equation can be written as
follows\begin{equation}i\hbar \frac{\partial \hat{O} (t)}{\partial t
} = \left [ \hat{O} (t), \hat{H} \right
],\end{equation}and\begin{equation}i\hbar \frac{\partial | \psi (t)
\rangle}{\partial t} = \hat{H} | \psi (t)
\rangle,\end{equation}respectively, where $\hat{H}$ is the
Hamiltonian of the system. As is well known, these two forms for the
equations of motion of quantum mechanics are equivalent
\cite{Dirac}. Of these, the Schr\"{o}dinger form seems to be the
more useful one for practical problems, as it provides differential
equations for wave functions, while Heisenberg's equation involves
as unknowns the operators forming the representative of the
dynamical variable, which are far more numerous and therefore more
difficult to evaluate than the Schr\"{o}dinger unknowns \cite{LL}.
Also, determining energy levels of various dynamic systems is an
important task in quantum mechanics, for this solving
Schr\"{o}dinger's wave equation is a usual way. Recently, Fan and Li
\cite{FL} showed that Heisenberg's equation can also be used to
deduce the energy level of some systems. By introducing the
conception of invariant `eigen-operator', they derive energy-level
gap formulas for some dynamic Hamiltonians. However, their
`invariant eigen-operator' equation involves operators as unknowns,
as similar to Heisenberg's equation, and hence is also difficult to
evaluate in general.

Recently, it is shown by the author \cite{Chen} that von
Neumann-Landau equation for wave functions is an extension of
Schr\"{o}dinger's wave equation and can be used to determine
energy-level gaps of the system. Contrary to Schr\"{o}dinger's wave
equation, vNLW is on `bipartite' wave functions. It is shown that
these `bipartite' wave functions satisfy all the basic properties of
Schr\"{o}dinger's wave functions which correspond to those
`bipartite' wave functions of product forms. Precisely, consider the
quantum system of a single particle. Note that the Hamiltonian for a
single particle in an external field
is\begin{equation}\hat{H}(\vec{x}) = - \frac{\hbar^2}{ 2 m}
\nabla^2_{\vec{x}} + U(\vec{x} ),\end{equation}where
$\nabla^2_{\vec{x}} = \partial^2/\partial x^2_1 +
\partial^2/\partial x^2_2 + \partial^2/\partial x^2_3,$
$U(\vec{x})$ is the potential energy of the particle in the external
field, and $\vec{x} = (x_1, x_2, x_3) \in \mathbf{R}^3.$ The
Schr\"{o}dinger's wave equation describing dynamics of the particle
is\begin{equation}i\hbar \frac{\partial \psi (\vec{x}, t) }{\partial
t} = \hat{H}(\vec{x}) \psi (\vec{x}, t) = - \frac{\hbar^2}{ 2 m}
\nabla^2_{\vec{x}} \psi (\vec{x}, t) + U(\vec{x} ) \psi (\vec{x},
t).\end{equation}The state of the particle can be described by a
definite wave function $\psi$ of Equation (1.4), whose stationary
states determine its energy levels. Moreover, the expectation value
of an observable $\hat{Q}$ in the state corresponding to $\psi$ is
determined by $\langle \hat{Q} \rangle_{\psi} = \langle \psi |
\hat{Q} | \psi\rangle.$

On the other hand, von Neumann-Landau equations for wave functions
are the following wave equations---von Neumann-Landau wave equations
(vNLW):\begin{equation}i\hbar \frac{\partial \Psi (\vec{x}, \vec{y};
t) }{\partial t} = \left (\hat{H}(\vec{x}) - \hat{H}(\vec{y}) \right
) \Psi (\vec{x}, \vec{y}; t),\end{equation}where $\Psi (\vec{x},
\vec{y}; t ) \in L^2_{\vec{x}, \vec{y}}.$ Contrary to
Schr\"{o}dinger's wave equation Eq.(1.4) for `one-partite' wave
functions $\psi (\vec{x}) \in L^2_{\vec{x}},$ the wave equation
Eq.(5) is an differential equation for `bipartite' wave functions
$\Psi (\vec{x}, \vec{y}),$ which, replacing $\hat{H}(\vec{x}) +
\hat{H}(\vec{y})$ by $\hat{H}(\vec{x}) - \hat{H}(\vec{y}),$ is also
different from Schr\"{o}dinger's wave equation for two particles.

We would like to mention that Eq.(1.5) has been presented by von
Neumann \cite{vN} and Landau \cite{LL} giving the change in the
density matrix with time. In \cite{Chen}, the author regard Eq.(1.5)
as a wave equation but not a equation for density functions. This is
the key point which is distinct from \cite{vN} and \cite{LL}. Then,
the author shows that the von Neumann-Landau wave equation Eq.(1.5)
is mathematically an extension of Schr\"{o}dinger's wave equation
Eq.(1.4) and a suitable form for motion of quantum mechanics as a
`bipartite' wave equation which presents a mathematical expression
of wave-particle duality.

Indeed, since$$\frac{\partial \left | \Psi (\vec{x}, \vec{y}; t)
\right |^2}{\partial t} = 2 \mathrm{Re} \left [ \Psi^* (\vec{x},
\vec{y}; t) \frac{\partial \Psi (\vec{x}, \vec{y}; t) }{\partial t}
\right ],$$it is concluded from Eq.(1.5) that\begin{equation}
\frac{\partial }{\partial t} \int \left | \Psi (\vec{x}, \vec{y}; t)
\right |^2 d^3\vec{x} d^3 \vec{y} = 0.\end{equation}This means that,
if this wave function $\Psi$ is given at some instant, its behavior
at all subsequent instants is determined. Hence, Schr\"{o}dinger's
wave equation is a special case of the wave equation Eq.(1.5) with
initial values of product form $\Psi (\vec{x}, \vec{y}; 0) = \psi
(\vec{x}) \psi^* (\vec{y}),$ because in this case $\Psi (\vec{x},
\vec{y}; t) = \psi (\vec{x}, t) \psi^* (\vec{y}, t)$ with $\psi
(\vec{x}, t)$ satisfying Eq.(1.4) and $\psi (\vec{x}, 0) = \psi
(\vec{x})$ and vice versa.

In this paper, we concern with the stationary von Neumann-Landau
wave equation. Similar to Laplace equation, we study the Dirichlet
problem for the stationary von Neumann-Landau wave
equation:\begin{equation}\left \{\begin{array}{l}\left ( -
\triangle_x + \triangle_y \right ) \Phi (x, y) = 0,~~~ x, y \in
\Omega,\\[0.4cm]
\Phi|_{\partial \Omega \times \partial \Omega} = F,
\end{array}\right.\end{equation} where $\Omega$ is a bounded domain in
$\mathbf{R}^n$ and $\triangle_x = \sum^n_{j=1}
\frac{\partial^2}{\partial x^2_j}$ for $x = (x_1, \ldots, x_n) \in
\mathbf{R}^n.$ Because the `bipartite' wave function $\Psi$ of
physical meaning should be `Hermitian':\begin{equation}
\Psi^*(\vec{x}, \vec{y} ) = \Psi(\vec{y}, \vec{x} ),\end{equation}it
is suitable to assume that\begin{equation}F \in C(\overline{\Omega}
\times \overline{\Omega}) \cap C^2(\Omega \times
\Omega),~~~\overline{F(x,y)} = F(y,x),\end{equation}for all $x, y
\in \overline{\Omega}.$ In this case, letting\begin{equation}W(x,y)
= \left ( \triangle_x - \triangle_y \right )
F(x,y),\end{equation}for all $x, y \in \Omega,$ we conclude that the
Dirichlet problem (1.7) reduces to\begin{equation}\left
\{\begin{array}{l}\left ( - \triangle_x + \triangle_y \right )
\Theta (x, y) = W(x,y),~~~ x, y \in
\Omega,\\[0.4cm]
\Theta|_{\partial \Omega \times \partial \Omega} = 0,
\end{array}\right.\end{equation}where $\Theta (x, y) = \Phi (x,y) - F(x,y).$

To solve (1.11) we will involve Hilbert-space's argument. Contrary
to Laplace equation, the ``energy functional'' of (1.11) is not
positive and so we shall utilize some additional techniques. To this
end, anti-inner product spaces are defined and studied in Section 2,
for which a Riesz-type representation theorem is proved. Finally, in
Section 3 we prove the existence and uniqueness of the generalized
solution of Dirichlet problem (1.11).

\section{Anti-inner product spaces}
In our terminology, an {\it inner product} on a vector space
$\mathbf{V}$ over the field $\mathrm{C}$ (or, $\mathrm{R}$) is a
complex-valued function $(\cdot, \cdot)$ defined for all pairs $x, y
\in \mathbf{V}$ so that the conditions$$( \alpha x + \beta y, z) =
\alpha (x,z) + \beta (y,z),~~\overline{(x, y)} = (y, x),$$are
fulfilled for all $x, y, z \in \mathbf{V}$ and $\alpha, \beta \in
\mathrm{C}$ (or, $\mathrm{R}$). If $(x, x) \geq 0$ for all $x \in
\mathbf{V}$ and $(x, x) = 0$ only if $x =0,$ the inner $(\cdot,
\cdot)$ is said to be definite; otherwise indefinite. A vector space
$\mathbf{V}$ over the field $\mathrm{C}$ (or, $\mathrm{R}$) equipped
with a definite (or, indefinite) inner product is called a complex
(or, real) inner (or, indefinite inner) product space (for details,
see \cite{Bognar}). It is well known that a complex (or, real) inner
product space $\mathbf{V}$ is a complex (or, real) normed vector
space under the norm $\| x \| = \sqrt{(x, x)},$ and said to be a
complex (or, real) Hilbert space if it is complete in this norm.

\

{\bf Definition 2.1}  An {\it anti-inner product} on a vector space
$\mathbf{V}$ over the field $\mathrm{C}$ (or, $\mathrm{R}$) is a
complex-valued function $\langle \cdot, \cdot \rangle$ defined for
all pairs $x, y \in \mathbf{V}$ so that the
conditions\begin{equation}\langle \alpha x + \beta y, z \rangle =
\alpha \langle x,z \rangle + \beta \langle y,z \rangle,~~\langle x,
y \rangle = -\langle y, x \rangle,\end{equation}are fulfilled for
all $x, y, z \in \mathbf{V}$ and $\alpha, \beta \in \mathrm{C}$ (or,
$\mathrm{R}$).

A vector space $\mathbf{V}$ over the field $\mathrm{C}$ (or,
$\mathrm{R}$) equipped with an anti-inner product is called a
complex (or, real) anti-inner product space.

An anti-inner product space $\mathbf{V}$ is called separated, if for
every $x \in \mathbf{V} \backslash \{0\}$ there is a $y \in
\mathbf{V}$ so that $\langle x,y \rangle \not=0.$

Throughout the paper, when not specified, we assume that all
anti-inner product spaces considered are separated.

\

{\bf Example 2.1} For $u, v \in \mathbf{C}^2,$
define\begin{equation}\langle u, v \rangle = \left |
\begin{array}{c}a ~~ b\\c ~~ d \end{array} \right |,\end{equation}where $u =
(a,b), v = (c,d).$ Then, $\mathbf{C}^2$ with (2.2) is a complex
anti-inner product space. Similarly, we can define an anti-inner
product on $\mathbf{R}^2$ so that $\mathbf{R}^2$ becomes a real
anti-inner product space.

{\bf Example 2.2} Let $\mathbf{V}$ be a complex inner (or,
indefinite inner) product space with the inner product $(\cdot,
\cdot).$ Define\begin{equation}\langle x, y \rangle = (x, y) - (y,
x) = 2i \mathrm{Im} (x, y)\end{equation}for all $x, y \in
\mathbf{V}.$ Then, $\mathbf{V}$ with (2.3) is a real anti-inner
product space. Here and afterwards, $\mathrm{Re} (a + i b) = a$ as
well as $\mathrm{Im} (a + i b) = b$ for all $a , b \in \mathrm{R}.$

Note that, in this case, $\mathbf{V}$ with (2.3) cannot be a complex
anti-inner product space. Indeed, by (2.3) $\langle x, y \rangle$ is
purely imaginary or zero. Suppose that $\mathbf{V}$ were a complex
anti-inner product space. Let $\langle x, y \rangle \not= 0.$ We
have$$\langle i x, i y \rangle = i \langle x, i y \rangle = -i
\langle iy, x \rangle = - \langle x,y \rangle \not= 0.$$ On the
other hand, $\langle x, i y \rangle$ is a purely imaginary number
and so $i \langle x, i y \rangle$ is a real number. Hence, $\langle
i x, i y \rangle = 0.$ This concludes a contradiction.

The anti-inner (2.3) is determined by the inner product of
$\mathbf{V},$ denoted by $\langle x, y \rangle_{\mathbf{V}}$ in the
following.

\

{\bf Definition 2.2}  An anti-inner product space $\mathbf{V}$ is
said to be complete, if there is a complex Hilbert space
$\mathbf{H}$ so that $\mathbf{V}$ is a real linear closed subspace
of $\mathbf{H}$ with the anti-inner product $\langle \cdot, \cdot
\rangle_{\mathbf{V}} = 2i \mathrm{Im}( \cdot, \cdot )_{\mathbf{H}}.$

A real linear functional on $\mathbf{V}$ is said to be bounded, if
it is bounded in the norm determined by $( \cdot, \cdot
)_{\mathbf{H}}.$

A linear functional $l$ on $\mathbf{V}$ is said to be {\it
representable}, if there is a $y \in \mathbf{V}$ so that
\begin{equation}l(x) = \langle x, y \rangle_{\mathbf{V}}\end{equation}for all $x \in
\mathbf{V}.$

\

{\bf Proposition 2.1}  Let $\mathbf{V}$ be an anti-inner product
space. Then
\begin{enumerate}
\item $\langle z, \alpha x + \beta y \rangle =
\alpha \langle z,x \rangle + \beta \langle z,y \rangle$ for all $x,
y, z \in \mathbf{V}$ and $\alpha, \beta \in \mathrm{C}$ (or,
$\mathrm{R}$).

\item For every $x \in \mathbf{V}, \langle x,x \rangle=0.$

\item Let $\mathbf{H}$ be a complex Hilbert space. Then, every real
closed subspace of $\mathbf{H}$ is a complete anti-inner product
space with the anti-inner product determined by $\langle \cdot,
\cdot \rangle_{\mathbf{H}}.$
\end{enumerate}

{\bf Proof}. By definition, the results are immediate.

\

{\bf Proposition 2.2} Let $\mathbf{V}$ be a complete anti-inner
product space determined by a complex Hilbert space $\mathbf{H}.$
Then, a bounded real linear functional $l$ on $\mathbf{V}$ is
representable if and only if $l$ is purely imaginary, that is,
$\mathrm{Re} l(x) =0$ for all $x \in \mathbf{V}.$

{\bf Proof}. By Definition 2.2, the necessity is evident. To prove
the sufficiency, we note that $\mathbf{V}$ is a closed subspace of
$\mathbf{H}$ and so a real Hilbert space. Then, by Riesz's theorem
there exists a $y \in \mathbf{V}$ such that$$l(x) = (x,
y)_{\mathbf{H}}$$for all $x \in \mathbf{V}.$ Since $l$ is purely
imaginary, it is concluded that $l(x) = i \mathrm{Im} (x, y)$ for
every $x \in \mathbf{V}.$ This completes the proof.

\section{The main result}
As is well known, a variety of methods have been invented to solve
the Dirichlet problem for the Laplace equation \cite{John}. Among
these is the method reducing the Dirichlet problem to a standard
problem in Hilbert space, that make little use of special features
of the Laplace equation, and can be extended to other problems and
other equations. In the sequel, we will generalize the Hilbert-space
method to solve the Dirichlet problem for the stationary von
Neumann-Landau equation. Typically the solution of the Dirichlet
problem proceeds in two steps by Hilbert-space method. In the first
step a modified (``generalized") Dirichlet problem is solved in a
deceptively simple manner. The second step consists in showing that
under suitable regularity assumptions on region and data the
``generalized" solution of the modified problem actually is a
solution of the original problem. The second step, which involves
more technical difficulties, will be carried out in another paper.
From the point of view of applications one might even take the
attitude that the modified problem already adequately describes the
physical situation.

We now reformulate the Dirichlet problem (1.7) for the stationary
von Neumann-Landau equation as the problem of representing a certain
bounded linear functional $l$ in a complete anti-inner product space
as an anti-inner product $\langle \Psi, \Phi\rangle.$ Under the
condition (1.9), it suffices to consider (1.11) instead of (1.7).

In the space of functions of class $C^1 (\overline{\Omega} \times
\overline{\Omega})$ we define two bilinear forms $( \Psi, \Phi )$
and $\langle \Psi, \Phi\rangle$ by\begin{equation}( \Psi, \Phi ) =
\int_{\Omega \times \Omega} \sum^n_{j=1} \frac{\partial
\Psi}{\partial x_j} \overline{\frac{\partial \Phi}{\partial x_j}} dx
dy + \int_{\Omega \times \Omega} \sum^n_{j=1} \frac{\partial
\Psi}{\partial y_j} \overline{\frac{\partial \Phi}{\partial y_j}} dx
dy\end{equation}and\begin{equation}\langle \Psi, \Phi\rangle =
\int_{\Omega \times \Omega} \sum^n_{j=1} \frac{\partial
\Psi}{\partial x_j} \overline{\frac{\partial \Phi}{\partial x_j}} dx
dy - \int_{\Omega \times \Omega} \sum^n_{j=1} \frac{\partial
\Psi}{\partial y_j} \overline{\frac{\partial \Phi}{\partial y_j}} dx
dy,\end{equation}respectively. Define $C^1_0 (\overline{\Omega}
\times \overline{\Omega})$ by\begin{equation}C^1_0
(\overline{\Omega} \times \overline{\Omega}) = \{ \Psi \in C^1
(\overline{\Omega} \times \overline{\Omega}): \Psi |_{\partial
\Omega \times \partial \Omega} = 0 \}.\end{equation} Then, $C^1_0
(\overline{\Omega} \times \overline{\Omega})$ with (3.1) is a
complex inner product space with the corresponding norm given by the
Dirichlet integral\begin{equation}\| \Psi \| = \sqrt{( \Psi, \Psi )}
= \left ( \int_{\Omega \times \Omega} \left [ \sum^n_{j=1} \left |
\frac{\partial \Psi}{\partial x_j} \right |^2
 + \sum^n_{j=1} \left |
\frac{\partial \Psi}{\partial y_j} \right |^2 \right ]dx dy \right
)^{1/2}.\end{equation}We complete $C^1_0 (\overline{\Omega} \times
\overline{\Omega})$ into a complex Hilbert space $H^1_0 (\Omega
\times \Omega)$ with respect to the Dirichlet norm (3.4).

On the other hand, define $S^1_0 (\Omega \times \Omega)$
by\begin{equation}S^1_0 (\Omega \times \Omega) = \{ \Psi \in H^1_0
(\Omega \times \Omega): \overline{\Psi(x,y)} = \Psi (y, x), a.e. (
x, y ) \in \Omega \times \Omega \}.\end{equation}Then, $S^1_0
(\Omega \times \Omega)$ is a real closed subspace of $H^1_0 (\Omega
\times \Omega).$ For $\Psi, \Phi \in S^1_0 (\Omega \times \Omega)$
define $( \Psi, \Phi )_S$ by\begin{equation}( \Psi, \Phi )_S =
\int_{\Omega \times \Omega} \sum^n_{j=1} \frac{\partial
\Psi}{\partial x_j} \overline{\frac{\partial \Phi}{\partial x_j}} dx
dy.\end{equation}By Fubini's theorem we have$$\begin{array}{lcl}(
\Psi, \Phi )_S & = & \int_{\Omega} \int_{\Omega} \sum^n_{j=1}
\frac{\partial \Psi (x,y)}{\partial x_j} \overline{\frac{\partial
\Phi (x,y)}{\partial
x_j}} dx dy\\[0.4cm]
& = & \overline{\int_{\Omega} \int_{\Omega} \sum^n_{j=1}
\frac{\partial \Psi (y,x)}{\partial x_j} \frac{\partial \overline{
\Phi (y, x) }}{\partial x_j} dx dy}\\[0.4cm]
& = & \overline{\int_{\Omega} \int_{\Omega} \sum^n_{j=1}
\frac{\partial \Psi (y,x)}{\partial x_j} \frac{\partial \overline{
\Phi (y, x) }}{\partial x_j} dy dx}\\[0.4cm]
& = & \overline{\int_{\Omega} \int_{\Omega} \sum^n_{j=1}
\frac{\partial \Psi (x,y)}{\partial y_j} \overline{\frac{\partial
\Phi (x, y) }{\partial y_j}} dx dy}.
\end{array}$$Hence, for $\Psi, \Phi \in S^1_0 (\Omega \times
\Omega)$ we have\begin{equation}( \Psi, \Phi ) = ( \Psi, \Phi )_S +
( \Phi, \Psi )_S\end{equation}and\begin{equation}\langle \Psi,
\Phi\rangle = ( \Psi, \Phi )_S - ( \Phi, \Psi )_S.\end{equation}In
particular,\begin{equation}\| \Psi \| = \sqrt{2} \| \Psi
\|_S,\end{equation}where $\| \Psi \|_S = \sqrt{( \Psi, \Psi )_S }.$
This means that $S^1_0 (\Omega \times \Omega)$ with (3.2) is a real
anti-inner product space. Since $S^1_0 (\Omega \times \Omega)$ is a
real closed subspace of $H^1_0 (\Omega \times \Omega),$ by the
Projection theorem we have that$$H^1_0 (\Omega \times \Omega) =
S^1_0 (\Omega \times \Omega) \oplus S^1_0 (\Omega \times
\Omega)^{\perp}.$$Set $\mathbf{H} = H^1_0 (\Omega \times \Omega)$
with the inner product $(\cdot, \cdot )_{\mathbf{H}}$ defined
by\begin{equation}(\Psi, \Phi )_{\mathbf{H}} = (\Psi_1, \Phi_1 )_S +
( \Psi_2, \Phi_2 )\end{equation}for $\Psi= \Psi_1 + \Psi_2, \Phi =
\Phi_1 + \Phi_2 \in \mathbf{H}$ with$$\Psi_1, \Phi_1 \in S^1_0
(\Omega \times \Omega),~~\Psi_2, \Phi_2 \in S^1_0 (\Omega \times
\Omega)^{\perp}.$$By (3.9) we have

\

{\bf Proposition 3.1} Set $\mathbf{V} = S^1_0 (\Omega \times
\Omega).$ Then, $\mathbf{V}$ with (3.2) is a complete anti-inner
product space determined by the Hilbert space $\mathbf{H}$ with the
inner product (3.10).

\

Let $\Theta \in C^2 (\overline{\Omega} \times \overline{\Omega} )$
be a solution of (1.11), where the prescribed $W$ belongs to $C
(\overline{\Omega} \times \overline{\Omega} ).$ Then, for any $\Psi
\in C^1_0 (\overline{\Omega} \times \overline{\Omega} )$ we have by
the divergence theorem\begin{equation}\begin{array}{lcl}\langle
\Psi, \Theta \rangle & = & \int_{\Omega \times \Omega} \Psi (x,y)
(-\Delta_x + \Delta_y) \overline{\Theta (x, y)} dx dy\\[0.4cm]
& = & \int_{\Omega \times \Omega} \Psi (x,y) \overline{W(x,y)} dx
dy.\end{array}\end{equation}This suggests that $\Theta$ can be found
by simply representing the known linear functional\begin{equation}l
( \Psi ) = \int_{\Omega \times \Omega} \Psi (x,y) \overline{W(x,y)}
dx dy\end{equation}as an anti-inner product $\langle \Psi, \Theta
\rangle.$ Our modified version (generalized solution) of the
Dirichlet problem (1.11) is then the following:

\

{\bf Definition 3.1} If $\Theta \in S^1_0 (\Omega \times \Omega)$
such that\begin{equation}\langle \Psi, \Theta \rangle = l ( \Psi )
\end{equation}for all $\Psi \in S^1_0 (\Omega \times \Omega),$ where
$\langle \Psi, \Theta \rangle$ and $l ( \Psi )$ are defined by (3.2)
and (3.12) respectively, then $\Theta$ is said to be the {\it
generalized solution} of the Dirichlet problem (1.11).

\

As following is the main result of the paper.

\

{\bf Theorem 3.1} Suppose $W \in L^2 (\Omega \times \Omega)$ so that
$\overline{W(x,y)} = - W(y,x)$ for almost all $(x,y) \in \Omega
\times \Omega.$ Then, there exists a unique generalized solution for
the Dirichlet problem (1.11).

{\bf Proof}. First, we have to show that the functional $l$ is
bounded. Since by the Cauchy-Schwartz inequality$$| l ( \Psi )|^2 =
\left | \int_{\Omega \times \Omega} \Psi (x,y) \overline{W(x,y)} dx
dy  \right |^2 \leq \int_{\Omega \times \Omega} |\Psi (x,y)|^2 dx dy
\int_{\Omega \times \Omega} |W (x,y)|^2 dx dy,$$and by
Poincar\'{e}'s inequality that there is a constant $M >0$ such that
$$\int_{\Omega \times \Omega} |\Psi (x,y)|^2 dx dy \leq M
\int_{\Omega \times \Omega} \left [ \sum^n_{j=1} \left |
\frac{\partial \Psi}{\partial x_j} \right |^2
 + \sum^n_{j=1} \left |
\frac{\partial \Psi}{\partial y_j} \right |^2 \right ] dx dy,$$for
all $\Psi \in H^1_0 (\Omega \times \Omega),$ we conclude by (3.9)
that $l$ is a bounded linear functional on $S^1_0 (\Omega \times
\Omega).$

Since $\overline{W(x,y)} = - W(y,x)$ for almost all $(x,y) \in
\Omega \times \Omega,$ we have by Fubini's theorem
$$\begin{array}{lcl}\overline{l ( \Psi )}& = & \int_{\Omega
\times \Omega} \overline{\Psi (x,y)} W(x,y) dx dy\\[0.4cm]
& = & - \int_{\Omega
\times \Omega} \Psi (y,x) \overline{W(y,x)} dx dy\\[0.4cm]
& = & - \int_{\Omega
\times \Omega} \Psi (y,x) \overline{W(y,x)} dy dx\\[0.4cm]
& = & - l ( \Psi )\end{array}$$for every $\Psi \in S^1_0 (\Omega
\times \Omega).$ This concludes that $l$ is purely imaginary.
Therefore, by Proposition 2.2 we prove the existence of the
generalized solution. The uniqueness of the generalized solution
follows immediately from the Riesz representation theorem. The proof
is complete.

\

For $f \in C^2(\overline{\Omega}),$ set $F(x,y) = f(x) f(y).$ Then,
$\Phi (x,y) = u(x) u(y)$ satisfies (1.7), where $u$ is the solution
of the Dirichlet problem for Laplace equation:\begin{equation}\left
\{\begin{array}{l} - \triangle u(x) = 0,~~~ x \in
\Omega,\\[0.4cm]
u|_{\partial \Omega} = f.
\end{array}\right.\end{equation}Generally speaking, $F$ cannot be
written as the product form of $F(x,y) = f(x) f(y),$ hence our
results extend the classical results for Laplace equation.

\ack{} This work was partially supported by the National Natural
Science Foundation of China under Grant No.10571176.

\end{document}